\documentstyle[11pt,newpasp,psfig]{article}
\begin{document}

\title{ Bar-driven Galaxy Evolution
and Time-scales to Feed AGN}
\author{F. Combes}
\affil{DEMIRM, Observatoire de Paris, 61 Av. de l'Observatoire,
F-75 014, Paris, France}

\begin{abstract}
 Recent progress in the understanding of the role of bars
and gravitational instabilities in galaxy disks is reviewed. 
It has been proposed that bars can produce mass transfer 
towards the center, and progressively metamorphose late-type
galaxies in early-types, along the Hubble sequence. Through
this mass transfer, bars are self-destroyed, and can act only
during a certain "duty-cycle" in the galaxy life. After sufficient
gas infall, another bar-phase can spontaneously occur. This 
recurrent evolution is strongly dependent on environment.
    A scenario is proposed, based on N-body simulations 
time-scales of the bar-life events, to explain the observed
bar frequency, gas mass fraction, bulge and possible black
hole mass growth, in a typical spiral.
\end{abstract}

\keywords{bars, dynamics, galaxy evolution, AGN}

\vspace*{-0.3cm}
\section{Formation and Evolution of Bars}

Bars as dynamical phenomena 
are much more understood than 20 years ago.
In the 70s, the first numerical simulations carried out
in the aim to find spiral structure, have established 
instead that bars are ubiquitous
(Miller et al. 1970, Hohl 1971, Hockney \& Brownrigg 1974).
Since then, numerous N-body simulations, considering 
only the stellar component, have confirmed that 
bars are robust, and stay for a Hubble time
(Sellwood 1981, Combes \& Sanders 1981).
But, as we shall see in this review, this is no longer true
when gas is taken into account! 

Bars can be considered as long-lived modes, as
superposition of leading and trailing waves
(or a standing wave). They grow through swing amplification
(e.g. Toomre 1981).  Waves can be 
reflected at the center, at corotation, but might be
damped at the inner Lindblad resonance (at least in the linear regime).
In N-body simulations,
the pattern speed takes first high values, larger than the peak 
of the $\Omega -\kappa$/2 curve,
then the bar slows down, and there exists one or two ILRs.

The bar instability begins in the center, where the 
precession speed of elliptical-like orbits, $\Omega -\kappa$/2,
 is high. Then, the bar traps more and more particles, which slows it down;
indeed, the particles farther from the center precess more slowly.
Angular momentum is taken away by the spiral, which 
amplifies the bar
(Sellwood \& Wilkinson 1993, Pfenniger \& Friedli 1991).

After their formation, 
bars can be further slowed down by dynamical friction 
against a dark matter halo, if it is concentrated enough
 (Debattista \& Sellwood 1998).
The fact that bars are not observed with low pattern speeds
seem to imply that the dark matter fraction inside
the bar radius is negligible, or that the dark matter
is rotating fast (Tremaine \& Ostriker 1999).

The existence of inner Lindblad resonances is
confirmed in many barred spiral observations: rings and 
characteristic resonant features are observed 
at ILRs (Buta \& Combes 1996). Also the gas
gas behavior, traced by dust lanes, shocks etc.. have confirmed
the predictions of simulations (Athanassoula 1992).

\vspace*{-0.1cm}
\section{Gas Instabilities, Regulation and Feedback}

The presence of gas changes considerably the picture.
Dissipation and star formation are essential phenomena.
The stellar component is heated by the spiral
waves and gravitational instabilities, and after
the bar is established, the system is 
then stable, in absence of gas. Torques
are exerted on the stars by the spiral wave
(since the potential and density extrema are not
in phase), but this is no longer the case for a bar,
by symmetry. A lenticular galaxy without gas can
stay barred for more than a Hubble time.

On the contrary, gas is continuously cooled, and spiral instabilities
are renewed, that influence the stellar component,
too  (Friedli \& Benz 1993, Heller \& Shlosman 1996).
Gravity torques from the bar to the disk matter
are therefore maintained, and no equilibrium is possible.
Since torques are continuously present, one can apply
the theory of ``viscous disks'': angular-momentum is
transferred outwards, to a small fraction of matter that
escapes to infinity, while the bulk of the disk mass is
driven inwards.
If the ``viscous'' time to redistribute angular momentum t$_{vis}$ 
is of the same order as the time scale to form  stars t$_*$, 
then an exponential disk of stars is created
 (Lin \& Pringle 1987a).

The radial redistribution of matter is done by
gravity torques. The viscosity here is therefore not the
normal gas viscosity (which anyhow is not efficient
in the interstellar medium at galactic scales), but 
the gravitational viscosity (Lin \& Pringle 1987b).
Gravitational instabilities are suppressed at small scales
by random motions, and the corresponding ``pressure'' 
through the local velocity dispersion $c$, and at large scale by rotation. 
The corresponding stability intervals are overlapping if
$Q = c \kappa /(\pi G \mu) = c/c_{crit} > 1$,
where $\mu$ is the disk surface density (Toomre 1964).
The evolution of the disk is then controlled by recurrent waves,
through a regulating mechanism, and feedback processes:

1) at a given epoch, the disk is cold (Q $<$ 1), 
and therefore unstable to spiral and bar waves

2) the disk then develops waves, non-axisymmetry and gravity 
torques, that transfer the angular momentum outwards (through 
trailing waves). The waves heat the disk, until Q is 1. 

3) then a disk with only a stellar component will remain stable, 
while a gaseous disk can cool back to step 1.

\noindent The size of the region over which angular momentum is tranferred is  
$\lambda_c \propto G \mu / \Omega^2$, and the corresponding 
time-scale is $2 \pi / \Omega$; the 
effective kinematic viscosity is   $\nu \sim \lambda_c^2  \Omega$ and the viscous 
time can be estimated as
  t$_{vis} \sim r^2/\nu \sim r^2 \Omega^3 / (G^2 \mu^2)$.
Now, there might be
approximate agreement between the two time-scales, viscosity and 
star-formation, since the two processes depend 
exactly on the same gravitational instabilities. 
Empirically, Q appears to control star-formation 
in spiral disks (Kennicutt 1989).
Then  t$_{vis}$ $\sim$ t$_*$, which could be the origin of
the exponential light distribution of the disk, and its
exponential metallicity gradient.
The time-scale for bar-driven evolution is of the order of a 
few dynamical time-scales. The controlling
time-scale is that of gas accretion, and depends 
strongly on the environment.

\begin{figure}[t]
\centerline{\psfig{figure=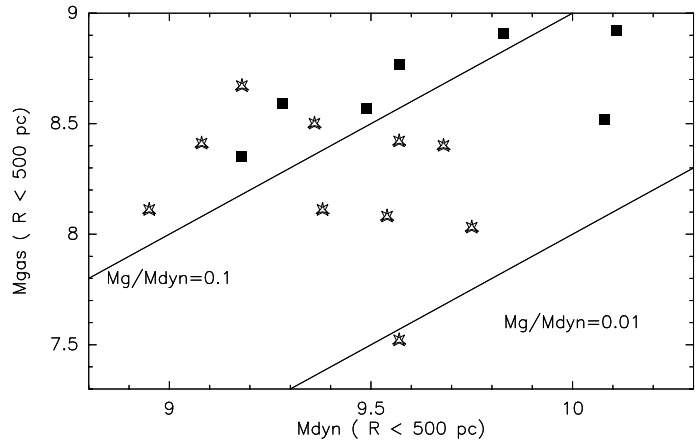,width=10cm,angle=-90}}
\caption{ Molecular gas and dynamical masses inside 500pc for
barred galaxies (filled squares) and non-barred galaxies (stars),
from Sakamoto et al. (1999)}
\label{sakam}
\end{figure}

That radial gas flows are efficient in barred galaxies is supported
by observations; barred galaxies have more H$_2$ gas concentration
inside their central 500pc than un-barred galaxies
(cf fig. \ref{sakam}, Sakamoto et al 1999). Also, the radial flows
level out abundance gradients in barred galaxies
(Martin \& Roy 1994).

\vspace*{-0.1cm}
\section{ Death of Bars and AGN Fueling}

\subsection{ Central Concentrations}

The inflow of matter in the center can destroy the bar.
It is sufficient that 5\% of the mass of the disk has sunk inside
the inner Lindblad resonance
(Hasan \& Norman 1990, Pfenniger \& Norman 1990, Hasan et al 1993).
But this depends on the mass distribution, on the size
of the central concentration; a point mass like a black hole is
 more efficient (may be 2\% is sufficient).
 The destruction is due to the mass re-organisation,
that perturbs all the orbital structure: the $x1$ orbits sustaining
the bar for instance are shifted outwards.
Near the center, the central mass axisymmetrizes
the potential. Then there is a chaotic region, 
and outside a regular one again.
 When a central mass concentration exists initially, in
N-body simulations, a bar still forms, but dissolves more
quickly. It is also possible that after a bar has dissolved,
another one forms, after sufficient gas accretion to 
generate new gravitational instabilities: the location 
of the resonances will not be the same.

If the radial inflow of gas is not violent, but slow enough,
the bar is weaken but not completely destroyed. The process
begins by the formation of two ILRs, through the mass
concentration. Since the periodic orbits inside the two ILRs
are perpendicular to the bars, this weakens the bar, and 
the mass flow is halted. The process is self-regulating
(Combes 1996; Sellwood \& Moore 1999).

A long term evolution of the bar is the box-peanut formation.
The stellar bar thickens through vertical resonances, in a Gyr time-scale
(Combes et al 1990, Raha et al. 1991). This does not destroy
the bar. But if the bar is destroyed afterwards by a central 
mass concentration, then this is a way to form
bulges. The observed 
correlation between scale-lengths of bulges and 
disks supports this mechanism (Courteau et al 1996).
Bars appear to exist in most box-peanut shape galaxies
(Merrifield \& Kuijken 1995, Bureau \& Freeman 1997).

\subsection{ Fueling the Nucleus}

If the mass concentration is not sufficiently  large,
the gravity torques accumulate the gas in a nuclear ring at ILR.
The curve $\Omega - \kappa$/2 (precession rate of elongated orbits in 
the epicylic approximation) is an increasing function of radius inside ILR.
On losing energy due to collisions and dissipation, the gas inflows,
and precesses then more slowly: it trails with respect to the
pattern. It experiences then positive gravity torques, and 
acquires angular momentum. It is therefore piling up back
into the ring.

If there exists a sufficient mass concentration (massive black hole),
the sense of  variation of  $\Omega - \kappa$/2 is reversed,
the gas leads, might form a leading spiral structure,
and experiences negative torques from the bar. 
The gas is driven further in, and can fuel the nucleus.
Fueling is possible, once a sufficiently massive  black hole 
is formed (Fukuda et al 1998).

\subsection{Bars within Bars}

When the mass accumulation grows in the center,
$\Omega - \kappa$/2 is increasing rapidly, 
and 2 ILRs are created with perpendicular $x2$ orbits.
Time-scales are therefore two different between the center
and outer parts, this forces the decoupling of a nuclear pattern
from the large-scale bar (Friedli \& Martinet 1993, Combes 1994).
Nuclear disks are frequently observed, 
kinematically as well (cf HST nuclear spirals, Barth et al 1995;
mm interferometers, Ishizuki et al 1990).

The second bars rotate with a much faster angular velocity.
To avoid chaos, the two bars have a resonance in common.
It is frequent that the ILR of the primary coincides with the 
corotation of the secondary. Multiply periodic particle orbits
have been identified in such time-varying potentials 
(Maciejewski \& Sparke 1998).
It is possible that the two bars exchange energy with each other,
through non-linear coupling; then $m=4$ and $m=0$ modes are
also expected, and these have been seen in simulations 
(Masset \& Tagger 1998).
Even then, the life-time of the ensemble is rather
short,  a few rotations. But the nuclear bars could help
to prolonge the action of the primary bar towards the
nucleus (as first proposed by Shlosman et al. 1989).

Other mechanisms are possible to help to fuel gas
into the nucleus; when the central nuclear disk becomes
gas-dominated, it is so unstable that clumps are formed,
creating a lot of non-axisymmetry (Heller \& Shlosman 1994).
Eventually, dynamical friction of giant molecular clouds on
the bulge, is very efficient, as soon as the clouds are inside
a couple hundred parsecs
(t$_{fric} \propto  r^2$ and is 3 10$^7$yr at r=200pc for a GMC 
of 10$^7$ M$_\odot$).

\subsection{Observational Tests of Bar-driven Fueling}

Observations have shown that dynamical perturbations
(bars or tides) are efficient to trigger nuclear starbursts
(Kennicutt et al 1987, Sanders et al 1988).
Besides, there are frequent associations between nuclear 
starbursts and AGN activity (Mirabel at al 1992).
Gas flow is the necessary condition to nuclear activity
and AGN-FRII (radio jets) are interacting
(Heckman et al 1986, Baum et al 1992), as well as
QSOs (Hutchings \& Neff 1992, Hutching \& Morris 1995).

But evidences for the correlation of bars and activity has
been difficult to obtain and controversial up to now.
The correlation is found in some samples
(Dahari 1984, Simkin et al 1980, Moles et al 1995),
while no more bars or interactions have been found for
Seyfert galaxies studied in the near-infrared 
(McLeod \& Rieke 1995, Regan \& Mulchaey 1999).

The case of Low Surface Brightness galaxies (LSB) is also
a puzzle. LSB are unevolved objects, in isolated environments,
with no mass concentrations. However 2/8 of them have Seyfert 
nuclei (instead of 1\% expected, Sprayberry et al 1995).
The probability of this configuration is only 2 10$^{-6}$.
Further work should be done here, to eliminate all
selection effects.

The lack of clear and obvious correlations between bars and
AGN might not be surprising, in view of the self-destruction
process described earlier: nuclear activity requires gas flows,
that are sufficient to destroy the bar. An anti-correlation
could even be expected, according to the chronology
and time-scales.

\begin{figure}[t]
\centerline{\psfig{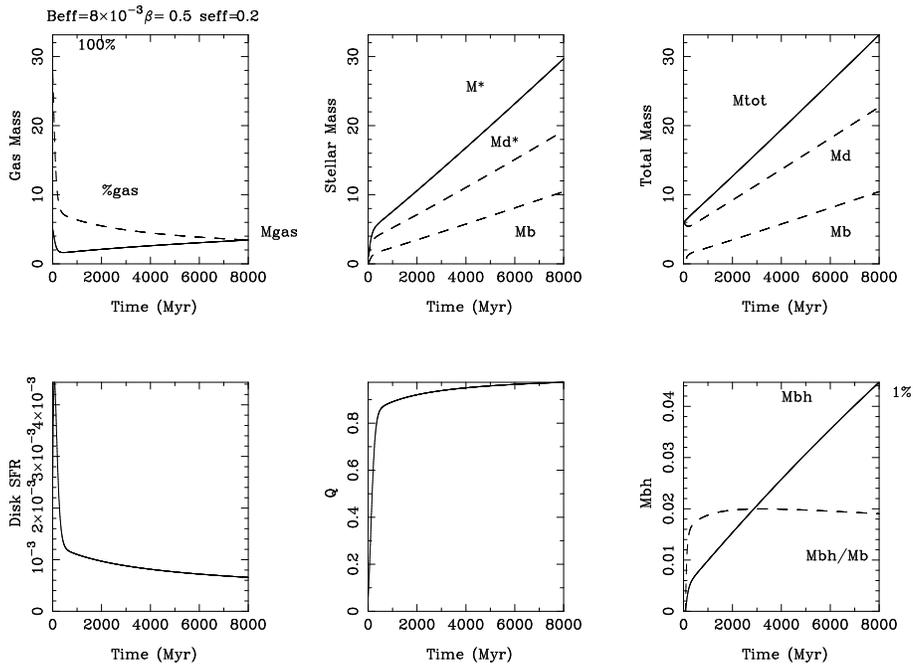}}
\caption{ Continuous gas accretion model: {\bf Top left} Full line: gas mass versus time;
dash line: gas mass fraction.  {\bf Top middle} Full line: stellar mass; dash lines: disk
stellar mass at top, and bottom bulge mass.   {\bf Top right} Full line: total mass;
dash lines: total disk mass at top, and bottom bulge mass. 
 {\bf Bottom left} Disk star formation rate versus time.  {\bf Bottom middle} 
Toomre Q parameter.  {\bf Bottom right}  Full line: mass of the central black hole,
and dash line: mass ratio between the black hole and the bulge.}
\label{evol_c}
\end{figure}

\vspace*{-0.1cm}
\section{Schematic Evolution Scenario}

Now that bar evolution is relatively well known, through N-body simulations,
consolidated by observations, it is interesting to test a toy model, in a
semi-analytical way, including:
\begin{itemize}

\item star formation, with a combination of
a quiescent rate, proportional to the gas 
density, in a time scale of 3 Gyr, and
a bar-driven contribution, with a threshold
(Q$<$1) and a rate equal to (1-Q )/t$_*$,
with t$_* = \beta$ t$_{vis}$.

\item radial flows: when a bar is formed, gravity torques
produce gas inflow, therefore with a threshold
Q$<$1 also,  and rate (1-Q )/t$_{vis}$,
with t$_{vis}$  $\sim \frac{1}{\Omega} (\frac{M_{tot}}{M_d})^2$.
There is also a radial flow of stars, with efficiency 
s$_{eff}$ = 0.2.

\item bulge formation: the inflowing gas (and stars) 
are assumed to form the bulge
through star-formation and vertical resonances

\item death of bars: when Q$>$1 
  (central concentrations, lack of gas and self-gravitating disk) 

\item gas infall: possibility of a continuous small infall  or a
periodically substantial  one (from companions).

\item black hole formation: a fixed fraction b$_{eff}$ of the radial 
gas flow is taken to contribute to its formation, i.e.
$ dM_{bh}/dt = b_{eff} M_g (1-Q)/t_{vis}$, with a threshold
Q$<$1.
\end{itemize}

\begin{figure}[t]
\centerline{\psfig{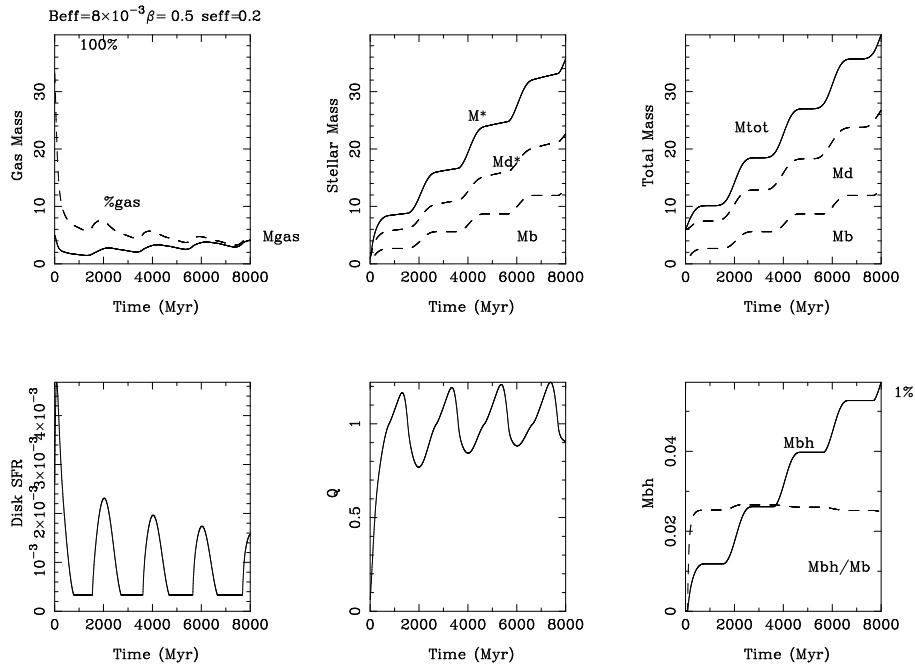}}
\caption{ Same as figure \ref{evol_c} for a periodic gas accretion} 
\label{evol_p}
\end{figure}

Figures \ref{evol_c} and \ref{evol_p} display some results of the toy 
model. The most striking feature is the self-regulation
of the stability parameter $Q$ towards 1. Although the 
galaxy initially starts almost completely gaseous, the gas
mass fraction soon stabilises to 10\% of the total. Also 
the mass of the central concentration (or black hole)
stabilises to a constant fraction of the bulge mass, as observed 
(Magorian et al. 1998).

\vspace*{-0.1cm}
\section{Conclusion}

Bars can be considered as one of the main
driver of disk evolution. They produce gravity
 torques that drive matter inwards. They are
gravitational instabilities that trigger
 star formation, with a time-scale comparable to that
of radial gas flows  (t$_*$ $\sim$ t$_{vis}$).
Bar formation is self-regulated when gas is present:  
massive infall, due to the bar, can weaken the bar.
This is accompanied by 
decoupling of a nuclear disk or second bar.
On a Gyr time-scale, stellar bars thicken,
through vertical resonsances, and the subsequent
destruction of the bar through gas flow 
leads to bulge formation.
 Bulges stabilise the disks and evolution is slower
with time, following bulge growing. The formation
of recurrent bars requires large gas infall.

Including these processes in a simple scenario, it
is easy to reproduce the following features:
tendency towards a nearly constant gas fraction 
(after a large decrease) and Q regulated around 1;
the secular building of the bulge; the
building and fueling of a central black hole.
The latter is increasing rapidly, from a
threshold in the bulge-to-disk mass ratio
of $\sim$  10\%.
Then the ratio of black hole mass to bulge mass 
tends to a constant,   of 0.5\%.
To get the observed order of magnitude for
this constant,  only 0.8\% of the gas flow towards the center
has to be taken to fuel the nucleus.

\end{document}